\documentclass[12pt]{article}
\usepackage{amsfonts}
\usepackage{amssymb}
\usepackage{latexsym}
\usepackage{graphicx}

\begin{document}

\begin{center}
{\Large \bf A small cosmological constant from the modified
Brans-Dicke theory -- an interplay between different energy scales} \\

\vspace{4mm}

Mikhail N. Smolyakov\\ \vspace{0.5cm} Skobeltsyn Institute of
Nuclear Physics, Moscow State University
\\ 119991
Moscow, Russia \\
\end{center}

\begin{abstract}
In this paper we discuss a model in which the energy density,
corresponding to the effective cosmological constant, after the
$SU(2)\times U(1)$ symmetry breaking appears to be of the desired
order of $10^{-48}\div 10^{-47} GeV^{4}$. The model contain two
different energy scales, one of which is associated with the
Higgs's vacuum expectation value. Another scale is of the order of
$10^{21}GeV$ and defines the vacuum expectation value of the
Brans-Dicke scalar field, non-minimally coupled to gravity, and
sets the value of the Planck mass. Other (dimensionless)
parameters are assumed not to contain hierarchical differences.
The model is devoid of any fine-tuning and gives a small value of
the effective cosmological constant even if the real "bare"
cosmological constant is quite large.
\end{abstract}

\section{Introduction}
During the last years the problem of cosmological constant, its
origin and small value attracts much attention. It is very likely
that the vacuum energy is indeed a constant, and there are a lot
of attempts to explain the existence of such constant vacuum
energy -- see, for example, reviews
\cite{Weinberg:1988cp}--\cite{Copeland:2006wr} and references
therein. One of the most interesting questions is that about its
extremely small value. Nevertheless, most of the mechanisms demand
a fine tuning, and corresponding cancellations do not look quite
natural.

In this paper we propose the model admitting any value of the real
cosmological constant and providing a desired value of the
effective cosmological constant. In the beginning, let us consider
a simple action describing two interacting scalar fields with the
action
\begin{equation}\label{actionsimple}
S=\int
d^{4}x\sqrt{-g}\left[-\frac{1}{2}\partial^{\mu}\phi\partial_{\mu}\phi-\frac{1}{2}\partial^{\mu}h\partial_{\mu}h-
\lambda_{1}\left(\phi^{2}-M^{2}\right)^{2}-\lambda_{2}\left(h^{2}-\frac{v^{2}}{2}\right)^{2}-\gamma\frac{h^{8}}{\phi^{4}}\right],
\end{equation}
where $\lambda_{1}\sim\lambda_{2}\sim 1$, $v\ll M$, the signature
of the metric is chosen to be $(-,+,+,+)$. The corresponding
vacuum solutions for the fields are given by the equations of
motion and look like
\begin{eqnarray}
\phi_{vac}\approx M+\frac{\gamma v^{8}}{32\lambda_{1} M^{7}},\\
h_{vac}\approx\frac{v}{\sqrt{2}}-\frac{\gamma
v^{5}}{4\sqrt{2}\lambda_{2} M^{4}}.
\end{eqnarray}
Let us suppose that $M=M_{GUT}\approx 10^{16} GeV$, $v=250 GeV$
and $\gamma=0.1$. The vacuum energy density
\begin{eqnarray}
\frac{\Lambda}{8\pi
G}=\lambda_{1}\left(\phi_{vac}^{2}-M^{2}\right)^{2}+\lambda_{2}\left(h_{vac}^{2}-\frac{v^{2}}{2}\right)^{2}+\gamma\frac{h_{vac}^{8}}{\phi_{vac}^{4}}
\end{eqnarray}
where $G$ is the gravitational constant, under these assumptions
takes the value
\begin{eqnarray}
\frac{\Lambda}{8\pi G}\approx \gamma\frac{v^{8}}{16
M_{GUT}^{4}}\sim 10^{-47} GeV^{4},
\end{eqnarray}
which is exactly the value of the observed dark energy density. In
some sense such a way of deriving the cosmological constant is
similar to obtaining the value of the vacuum energy by combining
the fundamental constants, -- such examples are discussed in
review \cite{Sahni:1999gb}. One can also recall the "seesaw"
mechanism for obtaining a small values of physical parameters
using very different energy scales, which has been recently used
in connection with the problem of cosmological constant
\cite{McGuigan:2006hs,Kiselev:2007bm,Kiselev:2007im}. At the same
time, at least for the action (\ref{actionsimple}), there can be
other contributions to the vacuum energy density, for example,
energy of the quantum fluctuations, which obviously neglects the
value obtained above. Nevertheless, the idea discussed in this
section can be used for constructing a model which gives the
necessary value of the effective cosmological constant even in the
case when the real vacuum energy density is much larger. We will
discuss this model in the next section.

\section{The model}\label{themodel}
Let us consider the action of the form
\begin{eqnarray}\label{gravact}
S=\int d^{4}x\sqrt{-g}\left[\alpha\phi
R-\omega\frac{\partial^{\mu}\phi\partial_{\mu}\phi}{\phi}-\lambda_{1}\left(\phi-M^{2}\right)^{2}-\bar\Lambda-
\frac{1}{48}F_{\mu\nu\rho\sigma}F^{\mu\nu\rho\sigma}-\right.\\
\nonumber\left.-\left(D_{\mu}H\right)^{\dagger}D^{\mu}H-
\lambda_{2}\left(H^{\dagger}H-\frac{v^{2}}{2}\right)^{2}+\gamma\left(H^{\dagger}H\right)^{2}\left(\frac{H^{\dagger}H}{\phi}\right)^{n}+L_{SM-Higgs}\right],
\end{eqnarray}
where $R$ is the four-dimensional curvature, $\phi$ is the
Brans-Dicke field, $\omega$ is the dimensionless Brans-Dicke
parameter, $H$ is the Higgs field, $\bar\Lambda>0$ is the "bare"
energy density of the vacuum and is supposed to include, for
example, contribution of quantum fluctuations, thus its value can
be large (in this action $\bar\Lambda=\rho_{vac}$ -- simply the
energy density of the vacuum, in this sense it is not the
cosmological constant defined by $\Lambda=8\pi G\rho_{vac}$),
$L_{SM-Higgs}$ is the Lagrangian of the Standard Model fields
without Higgs's kinetic term and potential. The Lagrangian of the
3-form gauge field is also added to the action to make the
cosmological constant in the Einstein equations be integration
constant (see \cite{Aurilia:1980xy,Henneaux:1984ji}). Constants
$\alpha$, $\gamma$, $\lambda_{1}$ and $\lambda_{2}$ are
dimensionless. We also suppose that $v\ll M$. Since we discuss a
theory which includes gravity, we do not take into account
possible issues concerning renormalizability of such theory. The
potential containing Higgs field can be also represented in
another form
\begin{eqnarray}
\left[\lambda_{2}-\gamma\left(\frac{H^{\dagger}H}{\phi}\right)^{n}\right]\left(H^{\dagger}H\right)^{2}-\lambda_{2}v^{2}H^{\dagger}H+\frac{\lambda_{2}v^{4}}{4}.
\end{eqnarray}
Of course, the constant term $\lambda_{2}v^{4}/4$ can be
incorporated into $\bar\Lambda$, but we retain it for simplicity.

Thus, we introduced explicit interaction between the Brans-Dicke
and Higgs fields. It should be noted that the idea to consider
interaction of the Higgs field with the fields related to gravity
was already discussed in the literature, see some examples in
papers \cite{vanderBij,higgscosm}. As for the use of the two
interacting scalar fields in cosmology, such constructions are
widely used in hybrid inflation models, see, for example, review
\cite{Linde:2007fr}. The Brans-Dicke field itself was also
discussed in cosmology \cite{GarciaBellido:1993wn}.

We suppose that the vacuum expectation value of the field $\phi$
is
\begin{eqnarray}\label{phivac}
\phi_{vac}&=&M^{2}.
\end{eqnarray}
Correspondingly, from equation of motion for the Higgs field we
get
\begin{eqnarray}\label{higgsvac}
\left(H^{\dagger}H\right)_{vac}&\approx
&\frac{v^{2}}{2}+\frac{\left(n+2\right)\gamma
v^{2n+2}}{2^{n+2}\lambda_{2} M^{2n}}.
\end{eqnarray}
Equation for the field $\phi$ gives us
\begin{equation}
\alpha
R=\gamma\left(H^{\dagger}H\right)^{n+2}\frac{n}{\phi_{vac}^{n+1}},
\end{equation}
which means that
\begin{equation}\label{BD-lambda}
R=\gamma\left(H^{\dagger}H\right)^{n+2}\frac{n}{\alpha\phi_{vac}^{n+1}}.
\end{equation}
Thus, the value of the effective cosmological constant is
\begin{equation}\label{lambdaeff}
\Lambda_{eff}=\frac{\gamma
n\left(H^{\dagger}H\right)_{vac}^{n+2}}{2\alpha\phi_{vac}^{n+1}}.
\end{equation}
The solution of equations of motion for the 3-form gauge field is
\begin{equation}\label{3formeq}
F^{\mu\nu\rho\sigma}\sim c\epsilon^{\mu\nu\rho\sigma},
\end{equation}
where $c$ is a constant, and contribution of this 3-form field to
the action reduces to
\begin{equation}
-\frac{1}{48}F_{\mu\nu\rho\sigma}F^{\mu\nu\rho\sigma}\to
+\frac{c^{2}}{2}.
\end{equation}
The constant $c$ in (\ref{3formeq}) is not fixed by the equations
of motion for the 3-form field. It is fixed by the Einstein
equations in accordance with (\ref{BD-lambda}). Indeed, it follows
from the contracted Einstein equations that
\begin{equation}\label{BD-lambda1}
\alpha\phi_{vac} R=2\left(\tilde\Lambda-\frac{c^{2}}{2}\right),
\end{equation}
where $\tilde\Lambda$ includes $\bar\Lambda$ and the constant
contributions coming from Higgs and Brans-Dicke potentials.
Finally
\begin{equation}\label{lambdaeff-c}
c^{2}=2\left(\tilde\Lambda-\alpha\phi_{vac}\Lambda_{eff}\right).
\end{equation}
Thus, this field makes the whole set of equations of motion
non-contradictory.

Now let us discuss possible energy scales of the model. The first
one is associated with the scale at which Higgs acquires its
vacuum expectation value. The second one, associated with $M$, can
be chosen by the following reasons. In the beginning of inflation
the gravity should be already "formed"\ at the classical level,
otherwise we would be unable to perform classical analysis of the
evolution. For the simplest model with the quartic inflaton
potential the initial value of inflaton field is roughly equal to
$10^{21}GeV$ \cite{Rubakov:2005tx}. We suppose that the field's
$\phi$ vacuum expectation value is of the same order and take
$M\sim 10^{21}GeV$. In this case its potential does not contribute
considerably to the energy density of the Universe, which appears
to be defined at that time only by the inflaton field. The vacuum
expectation value of $\phi$ also defines the value of the Planck
mass $M_{Pl}=\sqrt{\alpha\phi_{vac}}$.

Of course, there arises a question: why interaction of the
Brans-Dicke field with the Higgs has the form used in
(\ref{gravact})? Moreover, why these fields, one of which acquires
its vacuum expectation value before the beginning of the
inflation, interact one with another? We have no reasonable
answers to these questions and note that action (\ref{gravact})
should be interpreted only as a phenomenological model.

For $\alpha\sim 10^{-4}$, $n=3/2$ and $\gamma=1$ we get
\begin{equation}
\rho_{vac}^{eff}=\alpha\phi_{vac}\Lambda_{eff}\approx\frac{3\left(246^{2}/2\right)^{7/2}}{4(10^{63})}GeV^{4}\approx
3.6\cdot 10^{-48}GeV^{4}
\end{equation}
and
\begin{equation}
M_{Pl}=\sqrt{\alpha\phi_{vac}}\sim 10^{19} GeV.
\end{equation}

We would like to note that the choice of the energy scales made
above is not the only one possible. One can choose other values of
parameters of the model. Moreover, the gravitational constant can
be defined not by the Brans-Dicke field. Indeed, we can add the
following term to the action (\ref{gravact}):
\begin{eqnarray}
S_{extra}=M_{Pl}^{2}\int d^{4}x\sqrt{-g}R,
\end{eqnarray}
and suppose that the Brans-Dicke field is associated, for example,
with the energy scale of the symmetry breaking in a possible
theory of Grand Unification. In this case one can choose $M\sim
M_{GUT}\sim 10^{16}GeV$, which clearly shows that for $\alpha\sim
1$ the gravitational constant is mainly defined by $M_{Pl}$, not
by the Brans-Dicke field. This field is now simply the field
non-minimally interacting with gravity. Nevertheless, if one
chooses $\gamma\sim 1$, $\alpha\sim 1$, $n=2.3$, the necessary
value of $\Lambda_{eff}$ (see (\ref{lambdaeff})) also appears to
be obtained.

In the end of this section we would like to say a few words about
our choice of the vacuum expectation value for the field $\phi$.
The choice (\ref{phivac}) is not the only possible. One can take,
for example, a value slightly different from that used in
(\ref{phivac}). In this case the solution of corresponding
equations of motion would lead to another value of an effective
cosmological constant, as well as to another value of constant
$c$. Thus, $\phi_{vac}$ is a free parameter in some sense. At the
same time we suppose that Brans-Dicke field $\phi$ acquires its
vacuum expectation value (and defines the Planck mass) at very
high energies independently from the Higgs field (as well as from
other fields), and it is defined only by the term
$\lambda_{1}\left(\phi-M^{2}\right)^{2}$ of the potential (or by
an additional mechanism leading to (\ref{phivac})). This
assumption seems to be reasonable from the physical point of view.
When the Higgs field and the corresponding interaction with
Brans-Dicke field come to play at much lower energies, the Higgs's
vacuum expectation value appears to depend on $\phi_{vac}$.

It should be also noted that the Higgs field was used only because
its vacuum expectation value is quite convenient for obtaining the
necessary value of the cosmological constant. One can choose
another scalar field, even with a larger vacuum expectation value.
Anyway, a Brans-Dicke field (or another scalar field non-minimally
coupled to gravity) with a very large vacuum expectation value
should be used, since it "connects" different energy scales.

\section{Cosmological evolution}

Now let us discuss how the Brans-Dicke field can affect
cosmological evolution. First, we very briefly discuss the period
of inflation for the simplest case of a single inflaton field.
During the slow-roll regime the energy density of the universe is
supposed to be $V_{inf}\lesssim (10^{19}GeV)^{4}$. Let us consider
contracted Einstein equation
\begin{equation}\label{1inf}
M_{Pl}^{2}R\approx 2V_{inf}+2\lambda_{1}\varphi^{2},
\end{equation}
and equation for the Brans-Dicke field
\begin{equation}\label{2inf}
\alpha R\approx 2\lambda_{1}\varphi.
\end{equation}
Here $\phi=M^{2}+\varphi$, $\varphi\ll M^{2}$, $\lambda_{1}\sim
1$, we neglect the contribution of $\Lambda_{eff}$ and time
derivatives of the inflaton and Brans-Dicke field (indeed, during
the slow-roll period the curvature $R$  and inflaton field vary
slowly, thus the Brans-Dicke field also varies slowly as follows
from (\ref{2inf})). We also neglect contribution of the Higgs
field, which appears to be reasonable for such energy scale.

Multiplying (\ref{2inf}) by $M^{2}$ and combining with
(\ref{1inf}) we get
\begin{equation}
\varphi\approx\frac{V_{inf}}{\lambda_{1}M^{2}}\lesssim
10^{-8}M^{2}.
\end{equation}
We see that the assumption $\varphi\ll M^{2}$ is satisfied.

The contribution of the Brans-Dicke field to the energy density
can be approximated by the relation
\begin{equation}
\lambda_{1}\varphi^{2}\sim \frac{V_{inf}}{M^{4}}V_{inf}\lesssim
\frac{M_{Pl}^{4}}{M^{4}}V_{inf}\ll V_{inf}.
\end{equation}
We see, that the Brans-Dicke field does not make a significant
contribution to the energy density, and in this case inflation is
driven by the inflaton. Moreover, the Brans-Dicke field self-tunes
itself in an appropriate way to make the equation for the
Brans-Dicke field be satisfied.

Of course, corresponding analysis should be made much more
carefully, here we presented only rough reasonings.

Now let us turn to the evolution of the Universe at the present
time. Indeed, the evolution is governed not only by the
cosmological constant, but by ordinary and dark matter also. We
denote the (average) energy-momentum tensor of ordinary and dark
matter by $t_{\mu\nu}$.

Let us again consider contracted Einstein equation and equation
for the Brans-Dicke field with $\phi=M^{2}+\varphi$. We also
suppose that $\varphi\ll M^{2}$ and neglect time derivatives of
the Brans-Dicke field (indeed, Brans-Dicke's kinetic term takes
the form $\sim
\omega\frac{\partial^{\mu}\varphi\partial_{\mu}\varphi}{M^{2}}$,
which can be dropped in comparison with $\lambda_{1}\varphi^{2}$
for the evolution scale defined by $\Lambda_{eff}$). Below we will
show that these assumptions indeed are satisfied (at the same time
we should remember that actually field $\varphi$ depends on time
because of the time dependence of the trace of the energy-momentum
tensor $t=t_{\mu}^{\mu}$).

The corresponding contracted Einstein equation and equation for
the Brans-Dicke field look like
\begin{equation}\label{11inf}
M_{Pl}^{2}R\approx
2M_{Pl}^{2}\Lambda_{eff}+2\lambda_{1}\varphi^{2}-t.
\end{equation}
\begin{equation}\label{22inf}
M_{Pl}^{2}R\approx
2M_{Pl}^{2}\Lambda_{eff}+2\lambda_{1}M^{2}\varphi,
\end{equation}
from which it follows that
\begin{equation}
\varphi\approx -\frac{t}{2\lambda_{1}M^{2}}.
\end{equation}
We see that for the present average density of the Universe
$\varphi\ll M^{2}$, and our assumption is indeed satisfied. The
contribution of this field to the energy density
$$\lambda_{1}\varphi^{2}\sim\frac{t^{2}}{4\lambda_{1}M^{4}},$$
which is much smaller even than $M_{Pl}^{2}\Lambda_{eff}$ (we
realize that $t\sim M_{Pl}^{2}\Lambda_{eff}$), and thus can be
neglected. Again we see that the Brans-Dicke field does not affect
the evolution, its contribution to the energy-momentum tensor can
be dropped, and again it self-tunes itself in accordance with the
"ordinary" evolution governed by the Einstein equations.

In the end of this section it is necessary to discuss the problem
concerning the value of constant $c$. Indeed, it is defined in
accordance with equations (\ref{lambdaeff}) and
(\ref{lambdaeff-c}), which were obtained for the case of absence
of any matter except Higgs and Brans-Dicke fields in their vacuum
states. At the same time $c$ is a constant and does not depend on
time. In the limit $x^{0}\to\infty$ the ordinary and dark matter
average densities tend to zero, $\varphi\to 0$, solution for the
metric tends to $dS_{4}$ and only the cosmological constant
contributes to the energy density and pressure (the value of
$\bar\Lambda$ in (\ref{themodel}) is supposed to be that after all
the phase transitions such as electro-weak and QCD). Thus, the
constant $c$ should be such that equations of motion in this
asymptotic case be satisfied, i.e. it is defined by the boundary
conditions at the time infinity. This is exactly the value given
by (\ref{lambdaeff}).

We would like to note that in this section only a very brief
discussion of cosmological evolution is presented. One should make
a thorough analysis to make sure that the existence of the
Brans-Dicke field does not affect the cosmological evolution
significantly. But we suppose that since the parameter $M$ is very
large, the Brans-Dicke field would self-tune itself corresponding
to Einstein equations at all stages of the evolution, at least at
the classical level. At the same time the mechanism discussed in
Section~2 could "switch on" after the inflation period or even
later (for example, if the scalar field coupled to gravity is not
connected with definition of the Planck mass and acquires its
vacuum expectation value after this stage) and could not produce
any effect at these early stages. Anyway, at the present time,
when the energy/mass densities of usual baryonic matter, dark
matter and vacuum (defined by $\rho_{eff}$) are comparable, our
mechanism indeed works.

\section{Stability and Brans-Dicke--Higgs fields mixing}

In this section we will discuss how this model can modify Higgs
sector of the Standard model and its influence on the classical
Newtonian gravity. To this end we consider second variation
Lagrangian of the theory. Let us denote
$g_{\mu\nu}=g^{0}_{\mu\nu}+\frac{1}{M_{Pl}}h_{\mu\nu}$, where
$g^{0}_{\mu\nu}$ is the background metric, $\phi=M^{2}+\varphi$
and
\begin{equation}
H=\left(0\atop{\sqrt{(H^{\dagger}H)_{vac}}+\frac{\Phi}{\sqrt{2}}}\right),
\end{equation} then substitute it into action (\ref{gravact}) and
retain the terms quadratic in metric and scalar fields. We get
\begin{eqnarray}\label{gravactlin0}
S=\int
d^{4}x\sqrt{-g^{0}}\left[L_{2}[h_{\mu\nu}]-\frac{\omega}{M^{2}}
\partial_{\mu}\varphi\partial^{\mu}\varphi-\lambda_{1}\varphi^{2}-\frac{1}{2}\partial^{\mu}\Phi\partial_{\mu}\Phi
-\frac{m_{H}^{2}}{2}\Phi^{2}-\right.\\
\nonumber\left.-\frac{1}{M_{Pl}}h^{\mu\nu}\alpha\left(\varphi
R^{0}_{\mu\nu}-\varphi\frac{1}{2}g^{0}_{\mu\nu}R^{0}-\nabla_{\mu}\nabla_{\nu}\varphi+g^{0}_{\mu\nu}\nabla_{\rho}\nabla^{\rho}\varphi\right)-
\gamma\frac{n(n+2)\sqrt{2}\left(H^{\dagger}H\right)_{vac}^{n+\frac{3}{2}}}{M^{2n+2}}\Phi\varphi
+\right.\\
\nonumber\left.+\frac{1}{2M_{Pl}}h^{\mu\nu}g^{0}_{\mu\nu}\sqrt{2}\Phi\sqrt{\left(H^{\dagger}H\right)_{vac}}
\left(\frac{\gamma\left(n+2\right)\left(H^{\dagger}H\right)_{vac}^{n+1}}{M^{2n}}-2\lambda_{2}\left(\left(H^{\dagger}H\right)_{vac}-
\frac{v^{2}}{2}\right)\right)-\right.\\
\nonumber\left.-\frac{1}{2M_{Pl}}h^{\mu\nu}g^{0}_{\mu\nu}\varphi\frac{n\gamma
\left(H^{\dagger}H\right)_{vac}^{n+2}}{M^{2n+2}}+\frac{1}{2M_{Pl}}h^{\mu\nu}t_{\mu\nu}\right].
\end{eqnarray}
Here $L_{2}[h_{\mu\nu}]$ is the Lagrangian containing terms of the
second order in $h_{\mu\nu}$, $\nabla_{\mu}$ is the covariant
derivative with respect to the background metric $g^{0}_{\mu\nu}$,
$m_{H}$ is the mass of Higgs field $\Phi$ in the unitary gauge,
$R^{0}_{\mu\nu}$ and $R^{0}$ contain only $g^{0}_{\mu\nu}$,
$\omega\sim 1$, $\lambda_{1}\sim 1$ and other parameters are the
same as those used in Section~\ref{themodel}. We neglect
contributions $\sim\varphi^{2}$, $\sim\Phi^{2}$, coming from the
term describing Higgs--Brans-Dicke fields interaction, in
comparison with $\lambda_{1}\varphi^{2}$ and $m_{H}^{2}\Phi^{2}$
respectively. We also include interaction of the graviton with
matter with the energy-momentum tensor $t_{\mu\nu}$.

With the help of equations (\ref{higgsvac}), (\ref{BD-lambda}) for
the background solution this action takes a simpler form, where we
have also replaced $\left(H^{\dagger}H\right)_{vac}$ by
$\frac{v^{2}}{2}$ in the term $\sim\Phi\varphi$

\begin{eqnarray}\label{gravactlin01}
S=\int
d^{4}x\sqrt{-g^{0}}\left[L_{2}[h_{\mu\nu}]-\frac{\omega}{M^{2}}
\partial_{\mu}\varphi\partial^{\mu}\varphi-\lambda_{1}\varphi^{2}-\frac{1}{2}\partial^{\mu}\Phi\partial_{\mu}\Phi
-\frac{m_{H}^{2}}{2}\Phi^{2}-\right.\\
\nonumber\left.-\frac{1}{M_{Pl}}h^{\mu\nu}\alpha\left(\varphi
R^{0}_{\mu\nu}-\nabla_{\mu}\nabla_{\nu}\varphi+g^{0}_{\mu\nu}\nabla_{\rho}\nabla^{\rho}\varphi\right)-\gamma\frac{n(n+2)v^{2n+3}}{2^{n+1}M^{2n+2}}\Phi\varphi
+\frac{1}{2M_{Pl}}h^{\mu\nu}t_{\mu\nu}\right].
\end{eqnarray}
We see that some non-diagonal terms have vanished from the action.
For simplicity, we neglect the effect of the cosmological
constant. In this case $g^{0}_{\mu\nu}\to\eta_{\mu\nu}$, where
$\eta_{\mu\nu}$ is the flat Minkowski metric, $R^{0}_{\mu\nu}\to
0$ and $\nabla_{\mu}\to\partial_{\mu}$. In addition, let us make
redefinition $h_{\mu\nu}\Rightarrow
h_{\mu\nu}-\frac{\sqrt{\alpha}}{M}\eta_{\mu\nu}\varphi$.
Substituting it into action (\ref{gravactlin01}), we get
\begin{eqnarray}\label{gravactlin}
S=\int
d^{4}x\left[L_{FP}[h_{\mu\nu}]-\frac{1}{2}\left(\frac{2\omega+3\alpha}{M^{2}}\right)
\partial_{\mu}\varphi\partial^{\mu}\varphi-\lambda_{1}\varphi^{2}-\frac{1}{2}\partial^{\mu}\Phi\partial_{\mu}\Phi
-\frac{m_{H}^{2}}{2}\Phi^{2}-\right.\\
\nonumber\left.-\gamma\frac{n(n+2)v^{2n+3}}{2^{n+1}M^{2n+2}}\Phi\varphi
+\frac{1}{2M_{Pl}}h^{\mu\nu}t_{\mu\nu}-\frac{\sqrt{\alpha}}{2M_{Pl}M}\varphi\,t\right].
\end{eqnarray}
Here $L_{FP}$ is the standard Fierz-Pauli Lagrangian. After
redefining the fluctuations of the Brans-Dicke field as
$\tilde\varphi=\frac{\sqrt{2\omega+3\alpha}}{M}\varphi$,
(\ref{gravactlin}) takes the form
\begin{eqnarray}\label{gravactlin1}
S=\int d^{4}x\left[L_{FP}(h_{\mu\nu})-\frac{1}{2}
\partial_{\mu}\tilde\varphi\partial^{\mu}\tilde\varphi-\frac{m^{2}}{2}\tilde\varphi^{2}-\frac{1}{2}\partial^{\mu}\Phi\partial_{\mu}\Phi
-\frac{m_{H}^{2}}{2}\Phi^{2}-\right.\\
\nonumber\left.-\frac{b}{2}\Phi\tilde\varphi
+\frac{1}{2M_{Pl}}h^{\mu\nu}t_{\mu\nu}-\frac{\sqrt{\alpha}}{\sqrt{2\omega+3\alpha}}\frac{1}{2M_{Pl}}\tilde\varphi\,t\right].
\end{eqnarray}
Where $$m^{2}=\frac{2\lambda_{1}M^{2}}{2\omega+3\alpha}\sim
M^{2},$$
$$b=\frac{2\gamma
n(n+2)v^{2n+3}}{2^{n+1}\sqrt{2\omega+3\alpha}M^{2n+1}}\sim\frac{v^{6}}{M^{4}}.$$

The Lagrangian for the scalar fields can be easily diagonalized
with the help of the substitution
\begin{eqnarray}\label{diag1}
\Phi=\Phi'\cos{\theta}+\tilde\varphi'\sin{\theta},\\ \label{diag2}
\tilde\varphi=\tilde\varphi'\cos{\theta}-\Phi'\sin{\theta}
\end{eqnarray}
with $$\tan{2\theta}=\frac{b}{m^{2}-m_{H}^{2}}.$$ Since $\theta\ll
1$, we get
\begin{eqnarray}
\Phi=\Phi'+\frac{b}{2m^{2}}\tilde\varphi',\\
\tilde\varphi=\tilde\varphi'-\frac{b}{2m^{2}}\Phi'
\end{eqnarray}
and
\begin{eqnarray}
{m_{H}'}^{2}=m_{H}^{2}+m^{2}O\left(\frac{b^{2}}{m^{4}}\right)\approx m_{H}^{2},\\
{m'}^{2}=m^{2}+m^{2}O\left(\frac{b^{2}}{m^{4}}\right)\approx
m^{2}.
\end{eqnarray}
Here $$\frac{b}{m^{2}}\sim\frac{v^{6}}{M^{6}}\sim 10^{-112}.$$ The
action takes the form
\begin{eqnarray}\label{gravactlin2}
S=\int d^{4}x\left[L_{FP}(h_{\mu\nu})-\frac{1}{2}
\partial_{\mu}\tilde\varphi'\partial^{\mu}\tilde\varphi'-\frac{m^{2}}{2}\tilde{\varphi'}^{2}-\frac{1}{2}\partial^{\mu}\Phi'\partial_{\mu}\Phi'
-\frac{m_{H}^{2}}{2}{\Phi'}^{2}+\right.\\
\nonumber\left.+\frac{1}{2M_{Pl}}h^{\mu\nu}t_{\mu\nu}-\frac{\sqrt{\alpha}}{\sqrt{2\omega+3\alpha}}\frac{1}{2M_{Pl}}\tilde\varphi'\,t+
\frac{b\sqrt{\alpha}}{2m^{2}\sqrt{2\omega+3\alpha}}\frac{1}{2M_{Pl}}\Phi'\,t\right].
\end{eqnarray}
Quadratic action (\ref{gravactlin2}) can be also obtained in
another way. First, we can make a conformal rescaling in
(\ref{gravact}) and pass to the Einstein frame, then consider
quadratic approximation for the Lagrangian of the scalar fields
and fluctuations of the metric taking into account the equations
for the background metric, then make the diagonalization with the
help of (\ref{diag1}), (\ref{diag2}) and only finally pass to the
flat metric.

We see that the linearized theory does not contain tachyons or
ghosts, i.e. it is stable. The Brans-Dicke field and Higgs field
appear to be mixed, which leads to new interactions, for example,
of the Higgs field with matter through the trace of the
energy-momentum tensor $t$. But the corresponding new interactions
can be completely neglected because of the suppression by the
factor $10^{-112}$ for our choice of the parameters of the model.
We also see that the mass of the Brans-Dicke field $m\sim
10^{21}GeV$, so that this field decouples from the low-energy
effective theory. Thus, in the linear approximation, in fact, we
have ordinary tensor massless gravity.

\section{Conclusion}

In this paper we discussed a model which provides the necessary
value of the effective cosmological constant at the classical
level. We used interacting Higgs and Brans-Dicke fields and the
3-form gauge field. It is necessary to note that the Higgs and
Brans-Dicke fields are not the only possible fields which can be
used in the mechanism described above. The Higgs field was used
because of the value of the Standard Model's symmetry breaking
scale, while Brans-Dicke field was used because the corresponding
theory is one of the most known which can be used to define the
Planck mass within the framework of the classical field theory.
Obviously one can use any two interacting (in an appropriate way)
scalar fields with different vacuum expectation values, one of
which is non-minimally coupled to gravity, or even only one scalar
field, non-minimally coupled to gravity. But in the latter case
one should either introduce some energy scale "by hands" (in our
case it is provided by the Higgs mechanism), or use potentials
including, for example, exponential terms to make the hierarchy of
scales. Indeed, the idea is simple -- the effective cosmological
constant appears due to the interaction with Brans-Dicke field,
and the Higgs mechanism sets only the second energy scale. But
since we already have this scale, it is reasonable to use it, than
to introduce a new one. At the same time the 3-form gauge field is
necessary to make the "bare" cosmological constant be an
integration constant.

We think that it is also interesting to make a thorough study of
how the Brans-Dicke field in our model affects the whole
cosmological evolution. But this problem calls for further
investigations.

\section*{Acknowledgments}

The work was supported by grant of Russian Ministry of Education
and Science NS-8122.2006.2, by grant for young scientists
MK-8718.2006.2 of the President of Russian Federation, by grant of
the "Dynasty" Foundation and by scholarship for young teachers and
scientists of M.V.~Lomonosov Moscow State University. The author
is grateful to G.Yu.~Bogoslovsky, E.E.~Boos and especially to
I.P.~Volobuev for valuable discussions.

\end{document}